# Altering the reactivity of pristine, N- and P-doped graphene by strain engineering: a DFT view on energy related aspects

Ana S. Dobrota[a,*], Igor A. Pašti[a], Slavko V. Mentus[a,b], Börje Johansson[c,d,e], Natalia V. Skorodumova[c,d]

[a]*University of Belgrade – Faculty of Physical Chemistry, Belgrade, Serbia*

[b]*Serbian Academy of Sciences and Arts, Belgrade, Serbia*

[c]*Department of Materials Science and Engineering, School of Industrial Engineering and Management, KTH - Royal Institute of Technology, Stockholm, Sweden*

[d]*Department of Physics and Astronomy, Uppsala University, Uppsala, Sweden*

[e]*Humboldt University, Physics Department, Berlin, Germany*

**Abstract**

For carbon-based materials, in contrast to metal surfaces, the relationship between strain and reactivity is not yet established, even though there are literature reports on strained graphene. Knowledge of such relationships would be extremely beneficial for understanding the reactivity of graphene-based surfaces and finding optimisation strategies which would make these materials more suitable for targeted applications. Here we investigate the effects of compressive and tensile strain (up to ±5%) on the structure, electronic properties and the reactivity of pure, N-doped and P-doped graphene, using DFT calculations. We demonstrate the possibility of tuning the topology of the graphene surface by strain, as well as by the choice of the dopant atom. The reactivity of (doped) strained graphene is probed using H and Na as simple adsorbates of great practical importance. Strain can both enhance and weaken H and Na adsorption on (doped) graphene. In case of Na adsorption, a linear relationship is observed between the Na adsorption energy on P-doped graphene and the phosphorus charge. A linear relationship between the Na adsorption energy on flat graphene surfaces and strain is found. Based on the adsorption energies and electrical conductivity, potentially good candidates for hydrogen storage and sodium-ion battery electrodes are discussed.

**Keywords:** graphene, curvature, doping, reactivity

[*] corresponding author, email: ana.dobrota@ffh.bg.ac.rs

# 1. INTRODUCTION

Graphene, a monolayer of sp$^2$ hybridized carbon atoms arranged in a honeycomb lattice [1], is a zero-gap semimetal [2] chemically inert due to electronic delocalization. It has attracted significant scientific attention as a promising material for a wide range of applications [3–7] due to its many rare and unique properties [1,2,8]. When it comes to electrochemical energy conversion and storage, graphene-based materials are often proposed as the electrode materials in metal-ion batteries, fuel cells and supercapacitors [4,5,9,10], and are also discussed as potential materials for hydrogen storage [11–13]. The processes underlying the operation of the mentioned energy-related systems vary to a great extent - from electrocatalytic reactions and pseudo-faradaic processes to purely adsorptive processes. However, they all require the interaction of the electrode material with some chemical species of interest (*e.g.* metal ions, reactants and intermediates of electrocatalytic reactions, *etc.*). In case of electrocatalytic processes, the strength of the interaction of reactants and intermediates with the catalyst should not be too weak, nor too strong: it should be of an "optimal" strength [14]. If it is too weak, the reactants won't bind to the catalyst and the reaction won't take place; if it is too strong, the reactants, intermediates or products will act as catalytic poison. This leaves researchers with the task of figuring out what "optimal" means for each catalytic reaction of interest [15]. Pristine (ideal) graphene is not the best choice for the mentioned energy-related applications, precisely because of its chemical inertness - it interacts rather weakly with majority of chemical species. However, it is known in Materials Science that deviations from perfection can give rise to new accomplishments. Therefore, different strategies are developed for making graphene "less ideal" and more reactive towards the targeted species. These mostly include introducing defects to graphene by doping [16,17] and/or functionalizing it [18,19]. Another, less studied, way of tuning its properties is by deformation of its basal plane, *i.e.* introducing strain. Straining metal surfaces (either by epitaxial growth on a metal with different lattice constant, or by local deformation of a single metal phase) is known to modify the metal chemisorption properties considerably, for example, due to strain-induced shifts in the *d*-band of the metal [20]. This allows one to manipulate the metal reactivity using strain [20]. Such relationships are not established for carbon-based materials, including graphene, but would be extremely beneficial for understanding the reactivity of graphene-based surfaces and their rational modifications.

Tozzini and Pellegrini have theoretically demonstrated that the strength of H binding on graphene can be tuned by out-of-plane deformation, without introducing any surface functional groups [21]. Both theoretical and experimental studies have suggested that graphene sheets freely suspended in air or vacuum corrugate to some extent due to thermal fluctuations [22–24]. Corrugation of the graphene basal plane has also been observed for graphene supported on various substrates [25–27]. Additionally, the presence of certain functional groups can induce local corrugation of the graphene basal plane. It has been suggested that the dominant source of the enhanced reactivity of oxidized graphene compared to pristine graphene is the corrugation of the basal plane induced by binding of O-containing groups and sp$^2$ → sp$^3$ rehybridization [28]. According to π-orbital axis vector theory, the carbon atoms on highly curved surfaces should exhibit increased chemical potential due to diminished electronic delocalization, and consequently should be more reactive [29–31]. A simple empirical explanation of the enhanced reactivity of curved graphene was pointed out by Tozzini and Pellegrini [11]: convex surfaces have the sp$^2$ system distorted towards sp$^3$, which makes the protruding *p* orbitals more reactive and prone to covalent bonding (in their case, H binding). A more precise description of this effect, however, turned out to be quite complex, involving the interplay between mechanical distortion and electronic structure change due to plane deformation. Wu *et al.* have reported regions with higher degree of local curvature which demonstrate increased chemical reactivity towards aryl radicals on monolayer graphene grown by CVD on a Si-wafer substrate decorated with SiO$_2$ nanoparticles [32]. Boukhvalov and Katsnelson have predicted an enhancement in the reactivity of curved graphene if the height/radius ratio of the corrugation (ripple) is larger than 0.07 [33]. Okamoto *et al.* have found that H$_2$ physisorption energies on curved graphene clusters can be three times higher (absolute value) compared to those on planar graphene [34]. Curvature of the graphene basal plane was also offered as an explanation of enhanced H binding on microporous carbon, together with the presence of reactive edge sites in such systems [35]. Single walled carbon nanotubes and fullerenes can be considered as the "extreme" cases of graphene curvature. The hydrogen binding energies on their external surfaces is considerably more negative compared to that on flat graphene (by 1-2 eV per atom) [11]. Site-selective adsorption of atomic hydrogen on convexly curved regions of monolayer graphene grown on SiC(0001) was also reported using scanning tunneling microscopy [36]. However, the curvature affects graphene's reactivity in general, not only towards hydrogen [11]. On the other hand, stretching of the graphene basal plane (tensile strain) was reported to be beneficial for strengthening the bond between atomic oxygen and graphene. It increases the diffusion energy barrier of atomic oxygen on graphene, reduces the migration energy of oxygen through the graphene sheet, and shifts molecular oxygen dissociation from being endothermic to exothermic [37]. Stretching the graphene sheet by up to 10% has also been shown to stabilize metal clusters on graphene, strengthening the interaction by at least 100%, which is highly desirable for pursuing graphene-based catalysis [38]. Notice that graphene is known to be able to sustain reversible tensile elastic strain larger than 20% [39].

Here we investigate the influence of tensile and compressive strain on non-doped, as well as N- and P-doped graphene's structure, electronic structure and reactivity. We choose N and P as the representatives of n-type dopants in graphene which influence its structure in two ways: (i) the planar nature of graphene is not disturbed (N-doping), or (ii) significant plane deformation occurs (P-doping). To illustrate the reactivity of strained (doped) graphene surfaces we use H and Na as simple adsorbates, representatives of covalent and ionic bonding on graphene [40], which are also of great practical importance. Atomic hydrogen is an important adsorbate both from the aspect of various (electro)catalytic reactions and of hydrogen storage. Sodium, on the other hand, is one of the most abundant, non-toxic and low-cost elements. As it exhibits a high discharge voltage, it is seen as the heir of lithium in the world of metal-ion batteries. Due to larger diameter of Na$^+$ as compared to that of Li$^+$, its diffusion in graphitic materials is more difficult. Single-layered materials, such as graphene and modified graphene, offer improved Na$^+$ mobility and are expected to be good electrode materials for Na-ion batteries [41].

## 2. COMPUTATIONAL DETAILS

The results were obtained using the PWscf code of Quantum ESPRESSO distribution [42,43], which implemented ultra-soft pseudopotentials and plane waves basis set, within the generalized gradient approximation (GGA) and Perdew–Burke–Ernzerhof (PBE) exchange correlation functional [44]. Spin polarization was included in all the calculations. Dispersion interactions can make a great contribution to the overall interaction for the case of metal adsorption on graphene-based materials [40], so the correction was included through the DFT-D2 scheme of Grimme [45], as implemented in Quantum ESPRESSO [42,43]. Pristine graphene was modelled within a hexagonal 54 atoms supercell ($C_{54}$, ($3\sqrt{3} \times 3\sqrt{3}$)R30° structure in $xy$ plane), while doped graphene ($C_{53}X$, where X = N or P) surfaces were obtained by replacing one C with dopant atom (substitutional doping). The strain was introduced by compressing or elongating the supercell's $x$ and $y$ vectors by $n\%$, where $|n| \in \{1,3,5\}$. The models obtained by supercell compression are marked by negative strain percentage (e.g. $C_{53}N_{(-n\%)}$ is nitrogen-doped graphene compressed by $n\%$ along $x$ and $y$ cell vectors), while tensile strain is marked by positive percentage (e.g. $C_{53}P_{(+n\%)}$ is phosphorus-doped graphene stretched by $n\%$ along $x$ and $y$ cell vectors). The non-strained models are marked with 0%. The first irreducible Brillouin zone was sampled using a $\Gamma$-centered 4×4×1 grid of $k$-points generated by the general Monkhorst-Pack scheme [46]. The convergence with respect to the vacuum layer thickness and the $k$-point mesh was confirmed. Atomic positions were fully relaxed until the remaining forces acting on atoms dropped below 0.002 eV Å$^{-1}$. The deformation of the graphene layer was quantified in terms of the deformation energies ($E_{def}$), calculated as the difference in the total energies of the strained (compressed or stretched, $n\%$) and the corresponding non-strained (0%) system:

$$E_{def} = E_{C_{53}X_{(n\%)}} - E_{C_{53}X_{(0\%)}} \qquad (1)$$

where X = C, N or P.

The interaction of the chosen adsorbates (A = H or Na) with investigated surfaces was quantified in terms of their adsorption energies:

$$E_{ads}(A) = E_{subs+A} - (E_{subs} + E_{A,isol}) \qquad (2)$$

where $E_{subs+A}$, $E_{subs}$ and $E_{A,isol}$ stand for the ground state total energies of the substrate with A adsorbed, the total energy of the bare substrate, and the total energy of isolated A, respectively. By this definition, the molecular dissociation of $H_2$ is not included in $E_{ads}(H)$. The contribution of dispersion to the overall interaction of Na with graphene is relatively great, while in the case of H it is rather small [40]. Therefore, all the adsorption energies will be reported both from PBE (dispersion non-corrected) and PBE+D2 (Grimme's DFT-D2 dispersion correction scheme [45]) calculations. PBE+D2 results be given after PBE results, in parenthesis. A link between the adsorption energies obtained by PBE and PBE+D2 approaches will be discussed for both investigated adsorbates. Löwdin charges obtained with projwfc.x code of Quantum ESPRESSO [42,43] were used for charge analysis. Charge redistribution caused by sodium interaction with the chosen model systems was mapped by 3D charge difference plots, where charge difference ($\Delta\rho$) is defined as:

$$\Delta\rho = \rho_{subs+Na} - (\rho_{subs,frozen} + \rho_{Na}) \qquad (3)$$

where $\rho_{subs+Na}$, $\rho_{subs,frozen}$ and $\rho_{Na}$ stand for the ground state charge densities of the substrate interacting with Na, the substrate when Na is removed (frozen geometry), and that of isolated Na atom, respectively.

For conciseness, only the structures corresponding to 5% compression (−5%) and 5% stretch (+5%) will be presented in the figures. However, other investigated strain levels will also be discussed throughout the text.

## 3. RESULTS AND DISCUSSION

### 3.1. Curved (doped) graphene surfaces

Surfaces of non-doped, N-doped and P-doped graphene compressed or stretched by 1, 3 and 5% (as is described in Computational Details) were investigated, while the corresponding non-strained surfaces (0% strain) were used as reference points. For more details on the non-strained surfaces see ref. [47]. The topology of the resulting structures depends on the level of strain inflicted on the supercell, as well as on the presence of the dopant atoms. In general, negative strain, *i.e.* compression of the surface, results in plane corrugation, while for the positive strain the structure mimics the shape of the non-strained one, with some changes of the bond lengths.

Compressive strain generally resulted in two possible types of topology: "waves" and "hills" (examples can be found in **Fig. 1**). In the case of non-doped and N-doped graphene, the "waves" structure was found to be more stable for all investigated negative strain levels. In N-doped graphene, N preferred being located at the top or bottom of the wave (**Fig. 1**), rather than at its sides. P-doped graphene, however, exhibited some local curvature even in the no-strain case [47–49], due to larger atomic radius of P compared to C. Compressing such a structure only led to further pronouncing of the "hills" shape corresponding to the non-strained case, with P always being located at the hill top. Obviously, different topology of compressed graphene is obtained with different dopant atoms, *i.e.* the topology can be tuned by a specific combination of dopant atoms and strain.

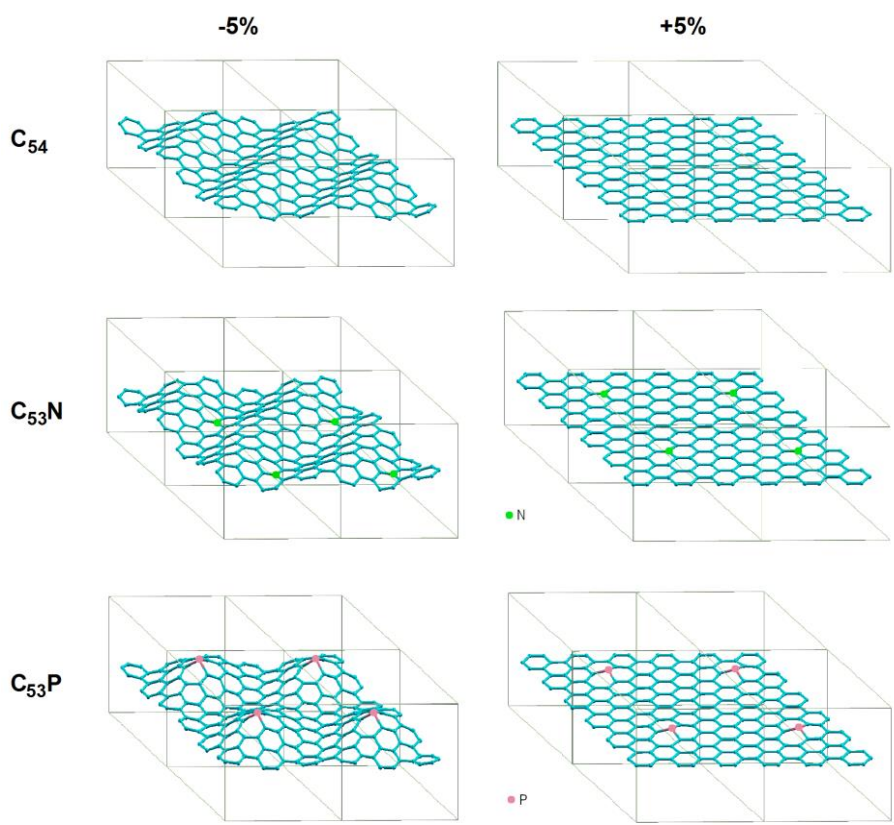

**Figure 1.** Relaxed structures of the most stable models of non-doped ($C_{54}$), N-doped ($C_{53}N$) and P-doped ($C_{53}P$) graphene compressed by 5% (−5%) and stretched by 5% (+5%). For easier comprehension of the overall topology, four adjoining supercells are shown. Gray lines represent the supercell borders. Graphical representation was made using XCrySDen [50].

Positive strain leads to stretching of the surface. Since non-doped and N-doped graphene are flat even in the no-strain case, they remain flat upon stretching, while C–C and C–N bonds become somewhat elongated (**Table S1**). On the other hand, the hill of P-graphene becomes less pronounced with rising tensile strain, and C–P bonds get shorter for strain up to +3%, but then elongate again going from +3% to +5% case (**Table S1**). The 3% compression cases (−3%) result in structures analogous to −5% case, but with somewhat less pronounced curvature. The structures corresponding to −1% strain are of the same shape as the non-strained (0%) cases, which are presented in ref. [47] in a smaller supercell.

Structures obtained with +1% and +3% strain are analogous to the 0% and 5% surfaces, with the P bump becoming less pronounced with growing tensile strain in P-doped graphene. Naturally, larger strain (absolute value) requires larger deformation energy (**Fig. S1**). Comparing the effects of equal compressions of pristine and doped graphene surfaces, it is clear that N- and P- doped graphene are easier to compress than pristine graphene. On the other hand, stretching the N-doped graphene requires more energy than non-doped and P-doped ones. Löwdin charges reveal that in the case of P-doped graphene the charge of the dopant atoms increases with more compressive strain, and decreases with rising tensile strain. In the case of N-doped graphene it shows the exactly opposite trend (**Table S1**).

All of the investigated non-doped and N-doped surfaces were found to be non-magnetic. When it comes to P-doped graphene, we find that the total magnetization is dependent on the strain level inflicted upon the supercell (**Table S1**). For the no-strain case, the total magnetization of P-graphene is found to be 0.97 $\mu_B$, similar to the available literature data [47,51,52]. For +3 and +5% cases, the total magnetization of P-graphene drops to zero. Thus, the magnetization of P-doped graphene can be controlled by strain. No band gap opening is observed upon graphene doping or straining (**Fig. 2**), *i.e.* the electrical conductivity is preserved. However, there is a significant alteration of the density of states compared to those of the non-strained cases.

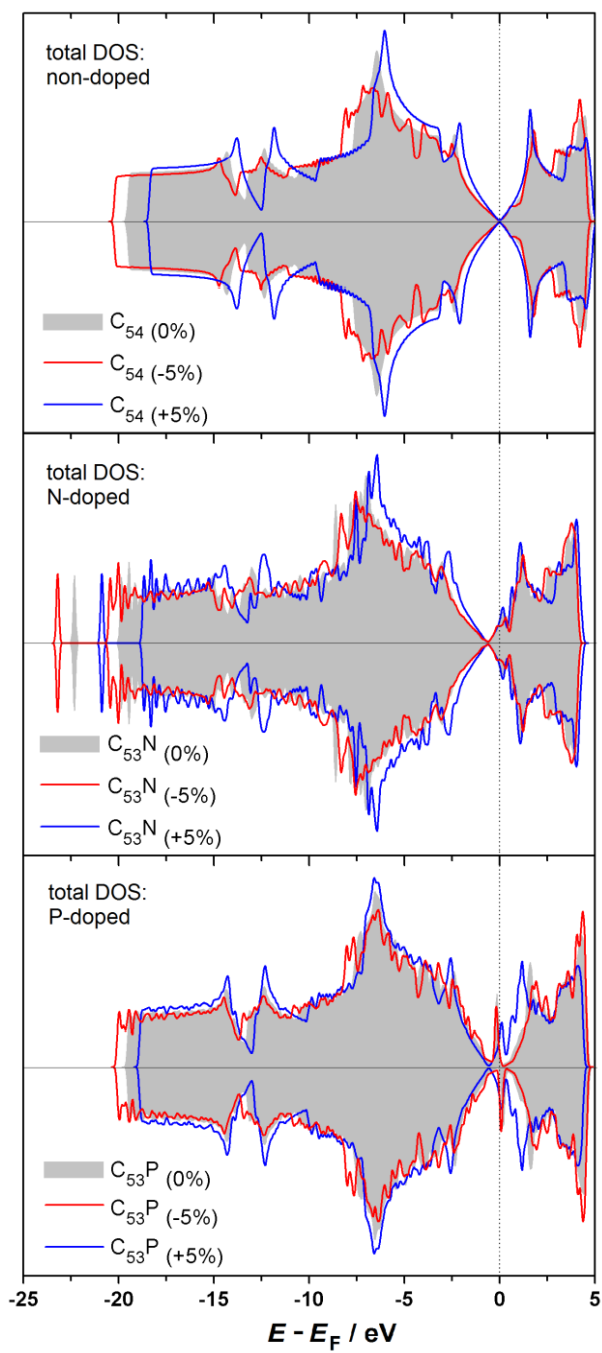

**Figure 2.** Total densities of state for 5% compressed (red) and 5% stretched (blue) surfaces of non-doped (top), N-doped (middle) and P-doped (bottom) graphene, given in comparison to the corresponding non-strained surfaces (grey shade).

It was previously shown that the corrugation of graphene basal plane can be significant for enhancing its reactivity [28]. In the continuation we focus on the reactivity of the investigated compressed and stretched (doped) graphene surfaces towards atomic hydrogen and sodium, as the representatives of covalent and ionic interactions with graphene, considering their importance in energy-related applications.

## 3.2. The reactivity of strained (doped) graphene

### 3.2.1. Atomic hydrogen adsorption

Atomic hydrogen forms covalent bonds with non-doped, non-strained graphene ($C_{54(0\%)}$), with preferential C-top adsorption site (~1.13 Å directly above one C atom) and the adsorption energy of −0.81 (−0.87) eV (in agreement with previous literature reports [28,40,53,54]). This interaction results in $sp^2 \rightarrow sp^3$ C rehybridization and a local deformation of the graphene basal plane: C atom bonded to H protrudes ~0.50 Å from the basal plane, while its first neighbours are shifted ~0.15 Å out of the plane. On non-strained N-graphene H preferentially binds to the dopant's first neighbouring C-top site, with the adsorption energy of −1.76 (−1.83) eV, and the corresponding C−H bond length of 1.12 Å. It was previously shown that carbon atoms in the nitrogen first coordination sphere are the most reactive ones towards H and OH in N-doped graphene [47]. When it comes to non-strained P-doped graphene, atomic hydrogen preferentially adsorbs to P-top site, *i.e.* on top of the dopant atom, with the adsorption energy of −2.50 (−2.55) eV, and the equilibrium P−H bond length of 1.445 Å, which is in agreement with previous report [47]). The preferential H adsorption site on the studied strained surfaces is the same as on the corresponding non-strained surfaces (Fig. 3).

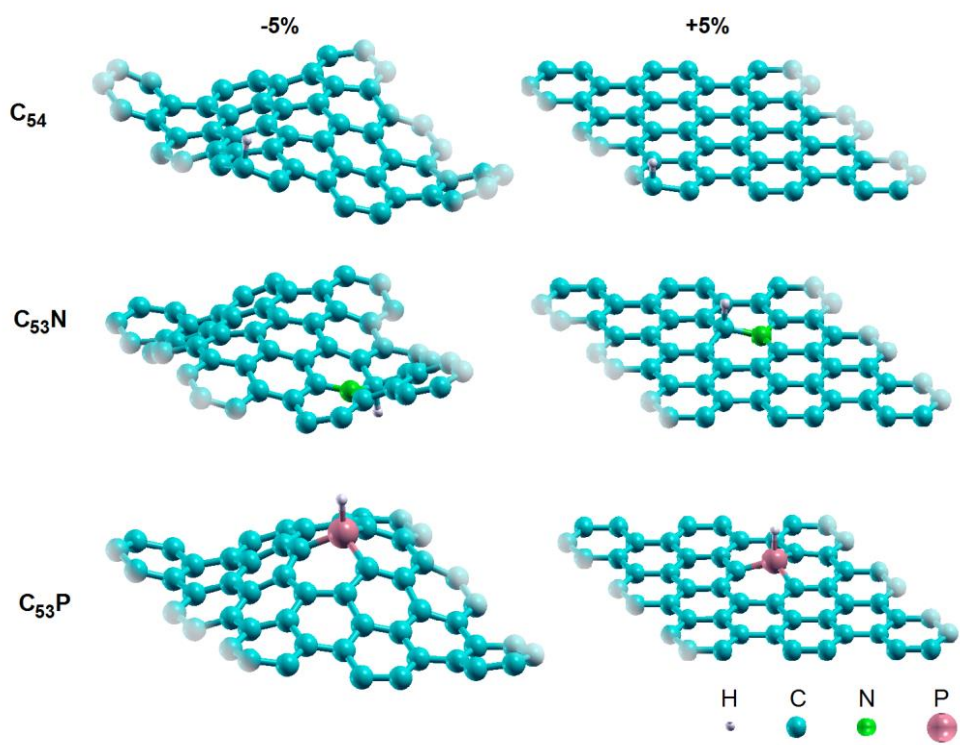

**Figure 3.** Relaxed structures of H adsorbed on non-doped (top row), N-doped (middle row) and P-doped (bottom row) graphene suppressed by 5% (−5%, left column) and stretched by 5% (+5%, right column). Graphical representation was made using XCrySDen [50].

For non-doped and N-doped graphene, all the investigated strain levels promote H bonding. However, the compressive strain strengthens this interaction more than the equal degree of tensile strain (**Fig. 4**). The enhancement of the reactivity of non-doped and N-doped graphene towards H with the corrugation of the plane is in agreement with our previous results which showed that the reactivity of the graphene basal plane increased with deformation, due to the formation of surface dangling bonds, which make C atoms in the deformed areas "bond-ready" [28]. Of course, in the case of doped graphene, the reactivity is also tuned by the charge redistribution induced by the presence of the dopant. Enhanced H binding on non-doped and N-doped graphene is in agreement with available literature reports [11,21,36], which qualify it for hydrogen storage applications.

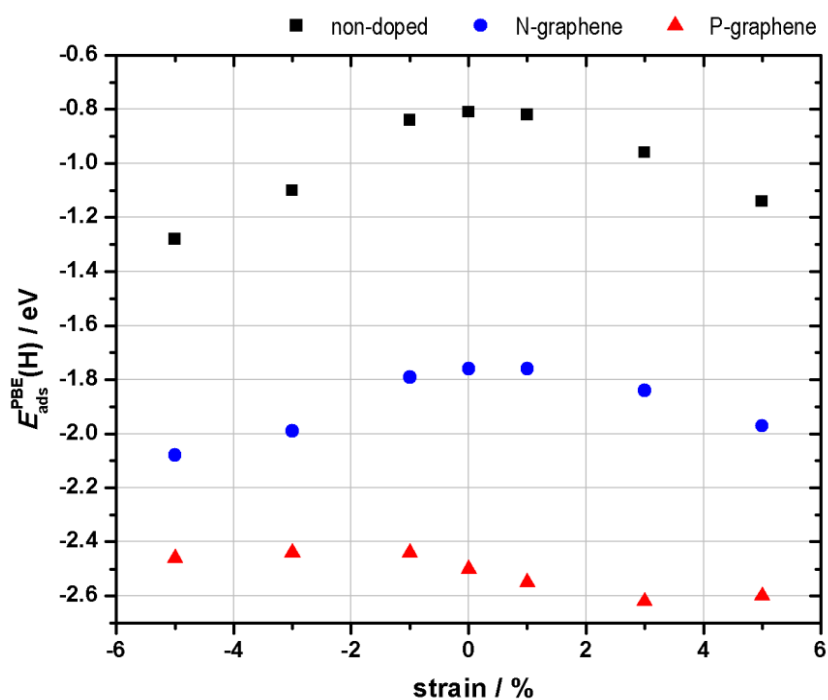

**Figure 4.** Adsorption energies of atomic hydrogen on pristine, N- and P- doped graphene under various strain levels, obtained by PBE calculations.

The case of P-doped graphene is somewhat different. As mentioned above, the most favourable H adsorption site in this case is P-top (**Fig. 3**). Löwdin charges indicate that H adsorption on this site results in some charge transfer from P to H atom. Upon compression of P-graphene hydrogen adsorption becomes slightly less exothermic. On the other hand, stretching P-graphene up to +3% results in strengthening of the H adsorption but it slightly weakens as tensile strain increases from +3% to +5% (**Fig. 4**). One can notice that the C−P bond length in bare $C_{53}P_{(n\%)}$ (**Table S1**) and the H adsorption energy on $C_{53}P_{(n\%)}$ (**Fig. 4**) follow similar trends when tensile strain is inflicted upon $C_{53}P$ ($n > 0$): while C−P distance is shortening, H adsorption is getting stronger, but its elongation is followed by weakening of H bonding (+5% case). Looking at the projected densities of states (**Fig. 5**), one can see that there is no band gap opening upon H adsorption and that the hydrogen's s state is well aligned with the binding atom's valence states (non-doped and N-doped graphene, those are carbon's p states, while for P-doped graphene that is phosphorus' p state, since H prefers binding to P-top site).

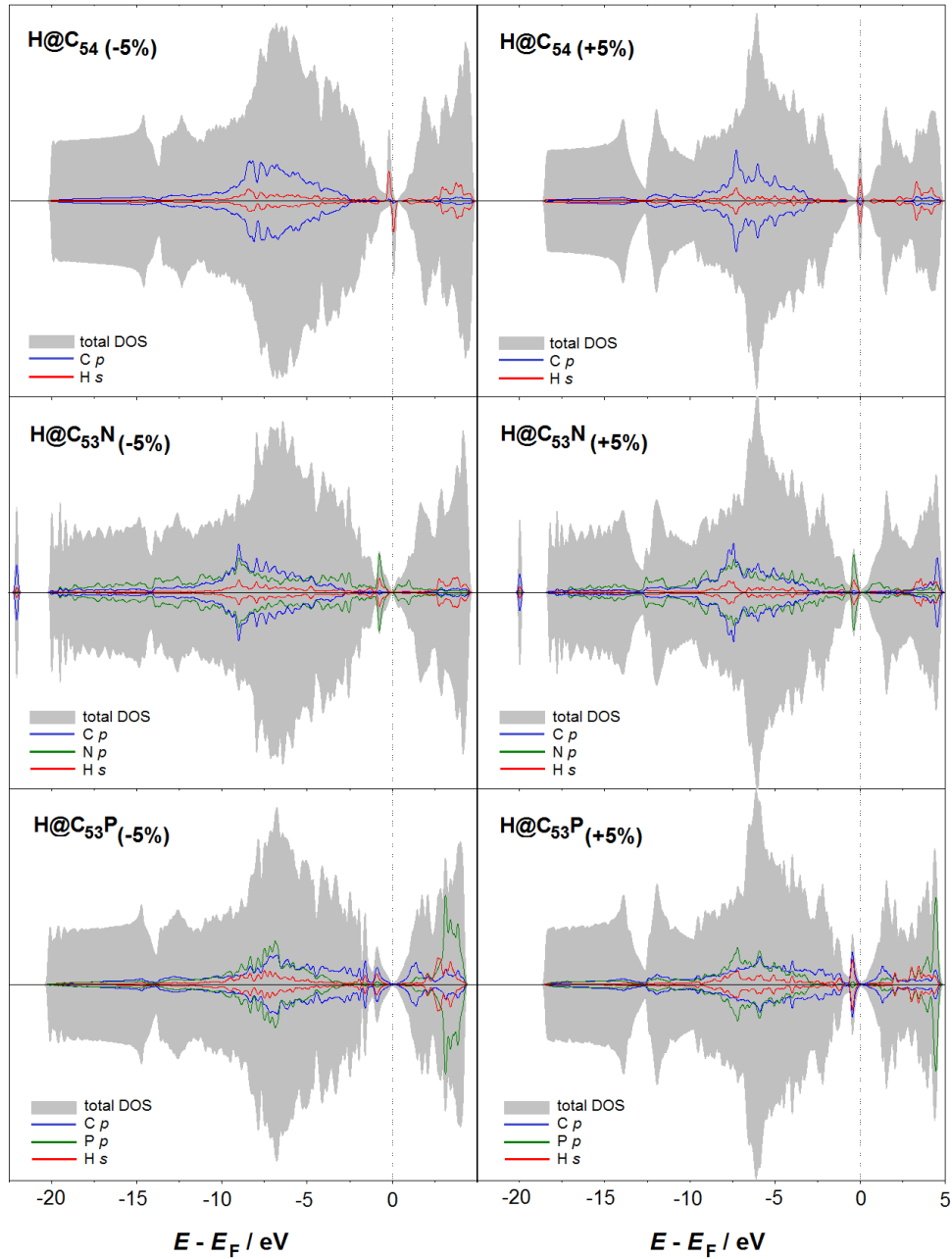

**Figure 5.** Electronic structures of H adsorbed non-doped (top row), N-doped (middle row) and P-doped (bottom row) graphene suppressed by 5% (−5%, left column) and stretched by 5% (+5%, right column). The Fermi level is set to 0 (vertical, dotted line). Partial DOS of H and the atoms which participate in the interaction are also provided.

### 3.2.2. Sodium adsorption

Sodium is known to interact rather weakly with pristine graphene, hovering more than 2 Å above its basal plane, with the preferential hollow adsorption site (above the centre of $C_6$ hexagon) [40,55–57]. Meanwhile, the graphene basal plane remains intact, as there is no structural change induced by Na adsorption. Here, we find that the corresponding Na adsorption energy amounts to −0.44 (−0.82) eV, in good agreement with previous literature reports [40,55–57], with some variations due to different supercell sizes and dispersion corrections used. Substitutional doping of graphene with N makes its interaction with Na even weaker [57,58], while P-doping strengthens it [57]. In the case of N-doped graphene, Na preferentially adsorbs at a $C_6$-hollow site as far as possible from the dopant, while in the case of P-doped graphene it prefers the site directly below the dopant, *i.e.* inside the bump made by P protrusion from the graphene plane [57]. We find the same here, with the corresponding Na adsorption energies of −0.30 (−0.66) eV and −1.31 (−1.75) eV for Na adsorption on N- and P-doped graphene, respectively, in good agreement with the mentioned reports.

Introducing strain can both strengthen and weaken the interaction of non-doped, N-doped and P-doped graphene with sodium, depending on the dopant and on the sign of the strain (compression or stretching), as it can be seen in **Fig. 6**. The cases of non-doped and N-doped graphene follow a similar trend: compressions up to 3% induce weakening of the interaction with Na, while 5% compression shifts the sodium adsorption energy back to the values similar to those for non-strained surfaces. On the other hand, stretching non-doped and N-doped graphene leads to monotonous strengthening of this interaction. In fact, we find a linear relationship between Na adsorption energy and the strain level of non-doped and N-doped graphene, for strains going from −1% to +5%. Since these strain levels correspond to the flat surfaces of non-doped and N-doped graphene, we can say that the Na adsorption energy is a linear function of strain as long as there are no structural deformations of the basal plane. We checked this by further stretching non-doped and N-doped graphene, up to +11% strain, and found that the linearity remained: the corresponding Na adsorption energies are −1.08 (−1.43) eV, -1.24 (−1.58) eV and −1.39 (−1.73) eV on non-doped graphene stretched by 7, 9 and 11%, respectively; and −0.95 (−1.31) eV, −1.12 (−1.47) eV, and −1.28 (−1.62) eV on N-doped graphene stretched by 7, 9 and 11%, respectively. P-doped graphene exhibits very different behaviour compared to this. Compression of P-graphene results in strengthening of Na binding, while stretching weakens it (**Fig. 6**). However, these adsorption energy changes are relatively small, and go up to ≈ 9%. According to PBE+D2 results, the strongest Na binding is found on the most curved $C_{53}P_{(-5\%)}$, with the adsorption energy of −1.24 (−1.80) eV, while PBE suggests strongest Na interaction with $C_{53}P_{(1\%)}$, with $E_{ads}$(Na) of −1.31 (−1.73) eV. Different behaviour of P-graphene compared to non-doped and N-doped graphene in this regard is somewhat expected, due to structural differences between them even in the non-strained case. Obviously, in the case of Na, including dispersion correction can affect overall conclusions.

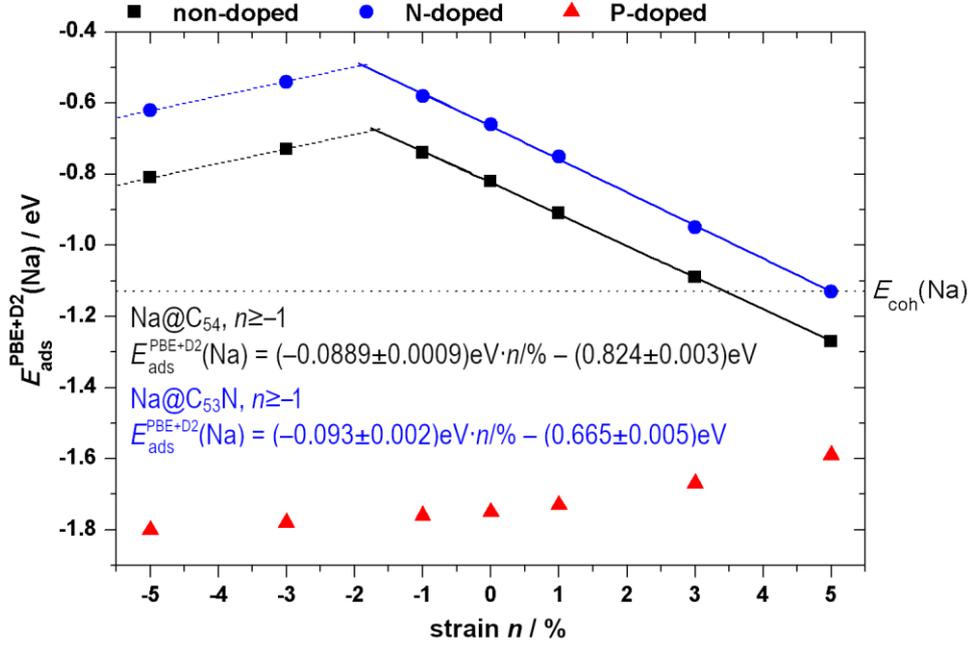

**Figure 6.** Adsorption energies of sodium on pristine, N- and P- doped graphene under various strain levels, obtained by PBE+D2 approach. Linear fits of data points for Na adsorption on flat (strain from −1% to +5%) non-doped and N-doped graphene, as well as associated regression equations are shown (colour coded). Sodium cohesive energy is marked by a dotted horizontal line.

The most stable structures of Na adsorption on −5% and +5% models are shown in **Fig. 7**. In general, strain does not change the preferential Na adsorption site. However, for non-doped and N-doped graphene surfaces, the hollow adsorption site is distorted for −3 and −5% compressions due to plane corrugation. In these cases, Na prefers settling inside of the wavy areas of these surfaces (**Fig. 7**). In the case of N-doped graphene, it prefers the dents which are far from the dopant. For P-doped graphene the preferential Na adsorption site is beneath the dopant, *i.e.* inside the bump made by P protrusion, for all investigated strain levels. In fact, in this case we observe a linear relationship between the Na adsorption energy and the Löwdin charge of the P atom in bare $C_{53}P_{(n\%)}$, without the adsorbate (**Fig. 8**). Going from −5% to +5% strain of $C_{53}P$, the charge of the dopant atom drops (**Table S1**) and the curvature diminishes (**Fig. 1**). In the same order, the distance between Na and P gets shorter upon adsorption, from 3.13 Å for Na@$C_{53}P_{(-5\%)}$ to 2.94 Å for Na@$C_{53}P_{(+5\%)}$. The strongest Na binding is observed on the most curved model, $C_{53}P_{(-5\%)}$, when the charge of phosphorus is the largest. No relationship analogous to the one in **Fig. 8** is observed for Na adsorption on non-doped or N-doped graphenes. In those cases Na is interacting with multiple atoms forming a hollow site, instead of direct interaction with just one (dopant) site.

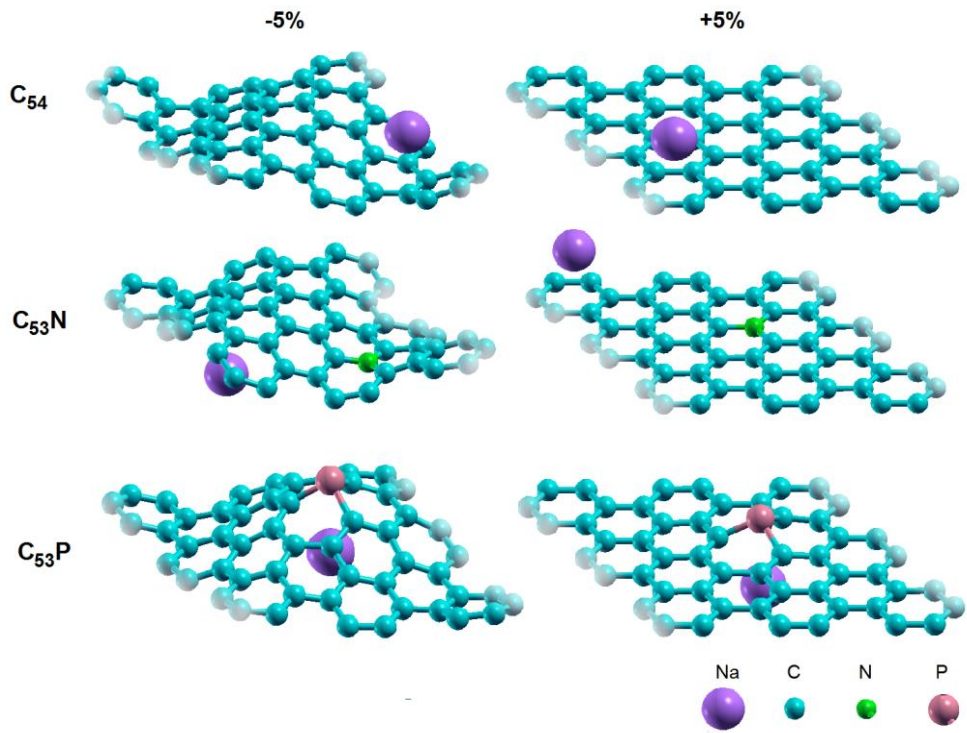

**Figure 7.** Relaxed structures of Na adsorbed on non-doped (top row), N-doped (middle row) and P-doped (bottom row) graphene suppressed by 5% (−5%, left column) and stretched by 5% (+5%, right column). Graphical representation was made using XCrySDen [50].

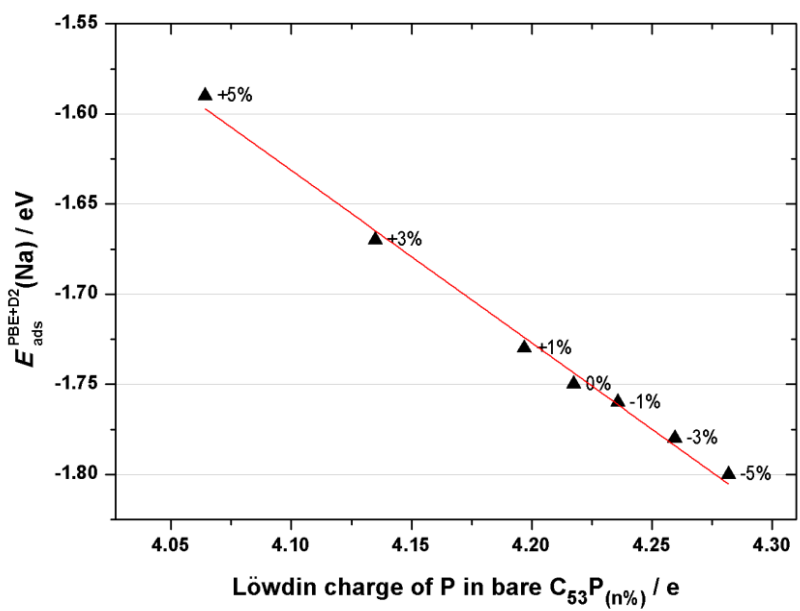

**Figure 8.** The relationship between sodium adsorption energies on P-doped graphene with various strain levels and the Löwdin charge of P atom in bare $C_{53}P_{(n\%)}$ (without the adsorbate). The corresponding strain level of $C_{53}P$ is given next to each data point.

The interaction between sodium and all the investigated systems is found to be dominantly ionic, as significant charge transfer from sodium to $C_{53}X_{(n\%)}$ is observed both from Löwdin charges and from charge difference plots (**Fig. 9**). In the cases of non-doped and N-doped graphene, charge from Na is transferred to the graphene basal plane. For the flat surfaces, it is transferred mostly to the carbon atoms which make up the preferential hollow adsorption site, while for curved surfaces the area where the charge settles is somewhat wider and encompasses more than one $C_6$ ring. When it comes to P-doped graphene, charge from Na is transferred to P as well as to the surrounding C atoms (**Fig. 9**). The ionic nature of Na bonding explains, to some extent, the relationship between phosphorus charge in bare $C_{53}P_{(n\%)}$ and $E_{ads}$(Na), shown in **Fig. 8**, as Na tends to transfer its electron to P and its neighbours. For all the cases, the Na s-states are located above the Fermi level and the total DOS (**Fig. 10**) looks much like the one of the same substrate before Na adsorption, except for the up-shift of the Fermi level due to the charge transfer from Na atom to the substrate.

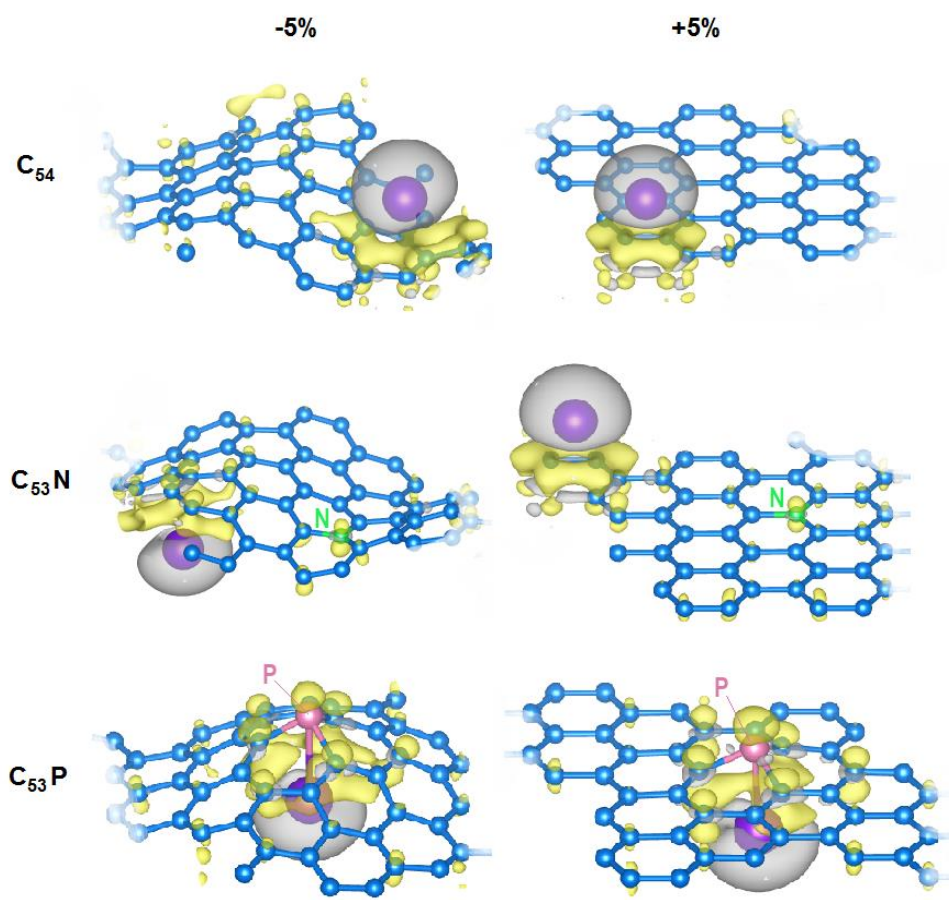

**Figure 9.** Charge difference plots for Na adsorption on non-doped (top row), N-doped (middle row) and P-doped (bottom row) graphene suppressed by 5% (−5%, left column) and stretched by 5% (+5%, right column). Yellow isosurfaces indicate charge gain, while grey isosurfaces indicate charge loss. Isosurface values are ±0.0015 e Å$^{-3}$. Graphical representation was made using VESTA [59].

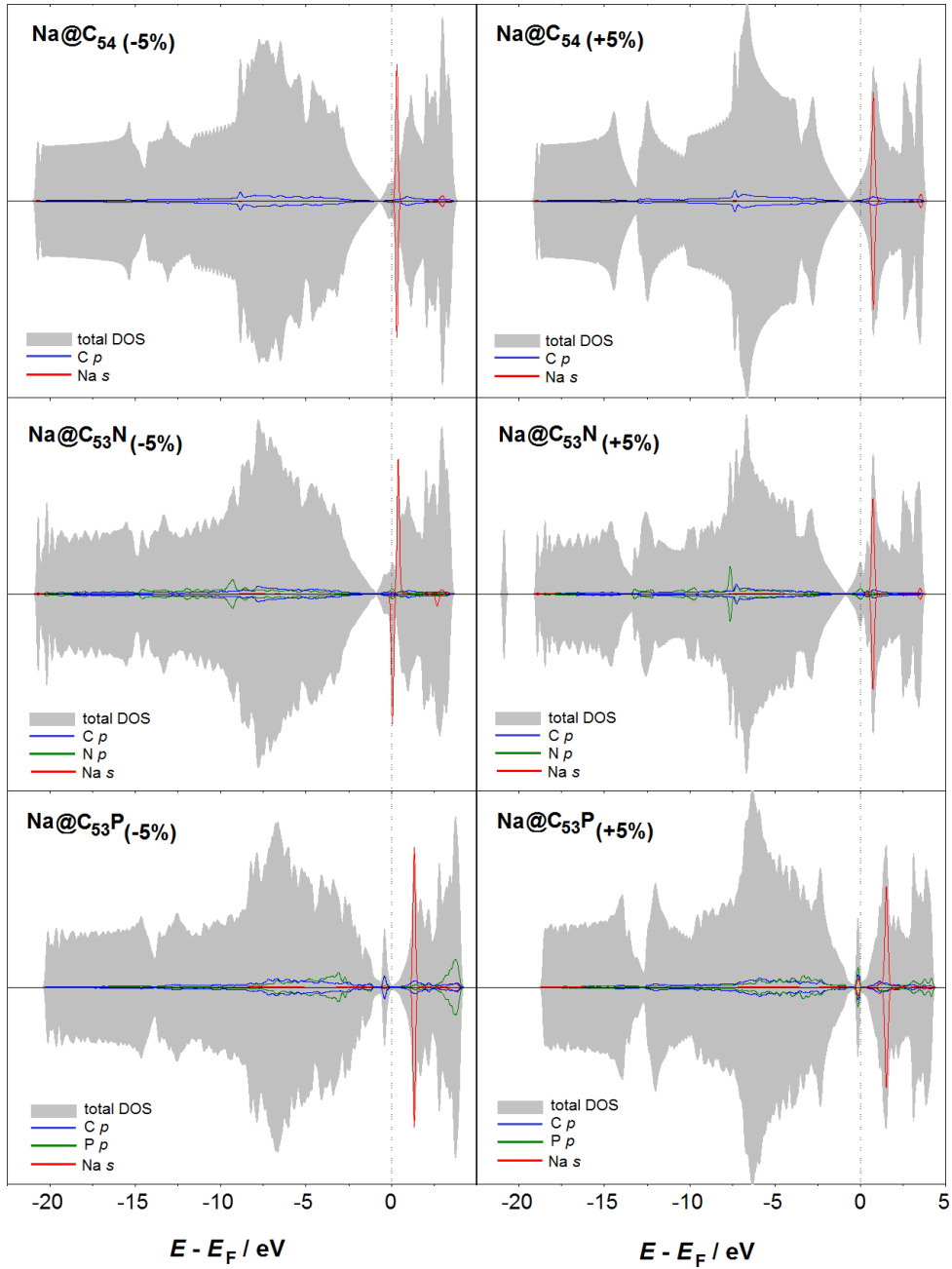

**Figure 10.** Electronic structures of non-doped (top row), N-doped (middle row) and P-doped (bottom row) graphene suppressed by 5% (−5%, left column) and stretched by 5% (+5%, right column), with Na adsorbed. The Fermi level is set to 0 (vertical, dotted line). Partial DOS of Na, the dopant (when there is one) and the atoms which participate in the interaction are also provided.

Sodium adsorption is of great practical importance when considering potential electrode materials for sodium-ion batteries. The cohesive energy of sodium (1.13 eV [60]) overcomes its interaction with pristine graphene, which could lead to metal phase precipitation during battery operation [61]. To avoid this, $E_{ads}$(Na) should be more negative than −1.13 eV. It is obvious from **Fig. 6** that this condition is fulfilled for +5% stretched non-doped graphene and all the investigated cases of P-doped graphene. Another important aspect in this regard is the electrical conductivity of the material. As shown in one of the previous sections the bare $C_{53}X_{(n\%)}$ systems are conductive (**Fig. 2**). DOS plots of Na@$C_{53}X_{(n\%)}$ systems reveal no band gap opening, *i.e.* the conductivity is preserved upon Na adsorption (**Fig. 6**). Based on Na adsorption energies and electrical conductivity considerations, the P-doped graphene is expected to be a very good candidate as a sodium-ion battery electrode material: no Na precipitation is expected, and it is conductive both before and after Na adsorption. However, it should be kept in mind that P is very reactive as a dopant in graphene, and not only towards Na. It is known to attract oxygen-containing groups to attach to it [47]. Therefore, for the potential application in Na-ion batteries, the environment should be strictly controlled if P is to be the active site towards Na. Non-doped graphene stretched by 5% is another potential candidate for the same application, with smaller "risk" of the active site for Na getting "poisoned" by another species.

### 3.2.3. Including dispersion correction

The dispersion-corrected H adsorption energies (PBE+D2) were found to scale extremely well with the corresponding dispersion-non-corrected (PBE) H adsorption energies (**Fig. 11**). A linear relationship between the H adsorption energies calculated by these two approaches is obtained, given by the formula: $E_{ads}^{PBE+D2}(H)=(0.989\pm0.004)E_{ads}^{PBE}(H)-(0.082\pm0.007)$ eV. Similar linear relationships have previously been observed for O and OH adsorption on pristine and B- and N- doped graphene surfaces [62]. This means that the inclusion of the dispersion correction does not alter the calculated chemisorption energies significantly, it just shifts them slightly more towards stronger binding (in our case, by ~ 80 meV).

When it comes to sodium, the dispersion-corrected adsorption energies (PBE+D2) are found to scale almost perfectly with the corresponding non-corrected (PBE) adsorption energies in case of no strain and positive strain (**Fig. 11**). However, the largest divergence from this linear relationship is observed for Na adsorption on the most corrugated surfaces: $C_{54(-5\%)}$, $C_{53}N_{(-5\%)}$, $C_{53}P_{(0\%)}$, $C_{53}P_{(-1\%)}$, $C_{53}P_{(-3\%)}$ and $C_{53}P_{(-5\%)}$. This indicates that in the case of Na adsorption on significantly corrugated surfaces the PBE+D2 adsorption energies cannot be easily estimated from the corresponding PBE results. When these data-points are excluded, a linear relationship between Na adsorption energies calculated by these two approaches is obtained: $E_{ads}^{PBE+D2}(Na)=(1.02\pm0.02)E_{ads}^{PBE}(Na)-(0.37\pm0.02)$ eV, in great agreement with the linear relationship obtained for the case of Na adsorption on doped (oxidized) graphene surfaces [57].

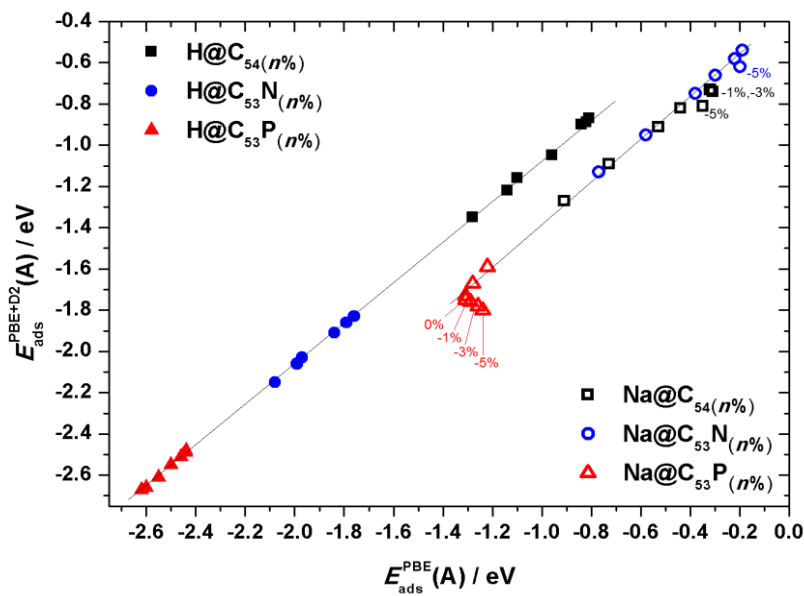

**Figure 11.** Hydrogen and sodium adsorption energies ($E_{ads}$(A), A = H or Na) on $C_{53}X_{(n\%)}$ systems, calculated using PBE+D2 *vs.* PBE approach. In case of Na adsorption, the corresponding strain level of $C_{53}X$ is given next to data points which deviate the most from the regression line.

## 4. CONCLUSIONS

We have investigated the effects of compressive and tensile strain (up to 5%) on the structure, electronic structure and the reactivity of non-doped, N-doped and P-doped graphene, by means of DFT calculations. The possibility of tuning the topology of the graphene surface by strain, as well as by the choice of the dopant atom, has been demonstrated. The reactivity of such systems has been probed using H and Na as simple adsorbates of great practical importance. Strain does not affect the preferential H adsorption site on (doped) graphene. In the case of sodium, the most favorable adsorption site can be affected by plane corrugation - sodium prefers settling in the dents on the surface. While H binds covalently to the investigated surfaces, Na exhibits an ionic interaction, with significant charge transfer from Na to graphene. Strain can either enhance or weaken the H and Na adsorption on (doped) graphene. In the case of Na adsorption, a linear relationship is observed between the Na adsorption energy on P-doped graphene and phosphorus' charge in corresponding bare substrate. For Na adsorption on flat graphene-based surfaces, a linear relationship between Na adsorption energy and strain is found - Na binds more strongly to more stretched surfaces. Surfaces of strained (doped) graphene are found to be conductive both in bare state (no adsorbate) and with H or Na adsorbed. Based on the adsorption energies of Na and electrical conductivity, P-doped graphene is expected to be a good candidate as a sodium-ion battery electrode material, but it should be kept in mind that P is very reactive in graphene, not only towards Na. Non-doped graphene stretched by 5% is another potential candidate for the same purpose, with less "risk" of the active site for Na adsorption to be "poisoned" by another species. The results obtained using PBE without any dispersion correction and those obtained with Grimme's D2 dispersion correction scale linearly with each other in the case of H adsorption, while there are some deviations from linearity in the case of Na adsorption on largely corrugated surfaces.

**Acknowledgements**

This work has been performed under the Project HPC-EUROPA3 (INFRAIA-2016-1-730897), with the support of the EC Research Innovation Action under the H2020 Programme, and was supported by the Serbian Ministry of Education, Science, and Technological Development (III45014). In particular, A.S.D. and N.V.S. gratefully acknowledge the support provided by PDC-KTH. S.V.M is indebted to Serbian Academy of Sciences and Arts for funding the study through the project ''Electrocatalysis in the contemporary processes of energy conversion''. N.V.S. acknowledges the support provided by Swedish Research Council through the project no. 2014-5993. We also acknowledge the support from Carl Tryggers Foundation for Scientific Research (grant no. 18:177), Sweden. The computations were performed on resources provided by the Swedish National Infrastructure for Computing (SNIC), at PDC-KTH and High Performance Computing Center North (HPC2N) at Umeå University.

# SUPPLEMENTARY INFORMATION

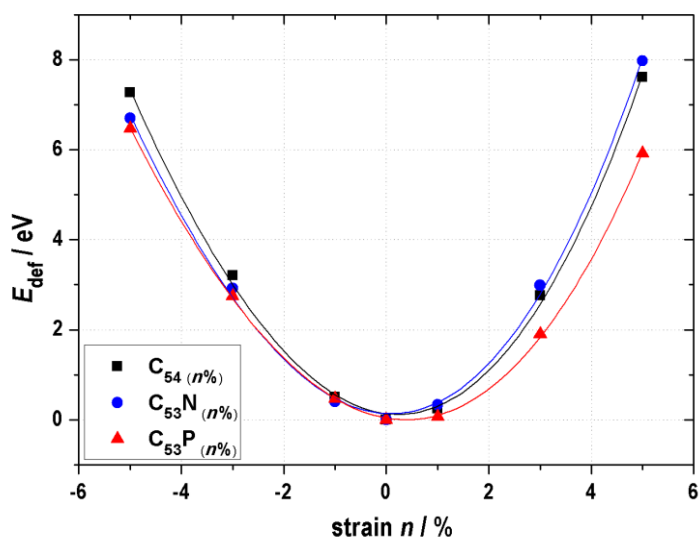

**Figure S1.** Deformation energies ($E_{def}$) of pristine and doped graphene surfaces, for various degrees of strain ($E_{def}$ is calculated with respect to the corresponding non-strained (0%) surfaces).

**Table S1.** Average C−X bond lengths and Löwdin charges of dopant atoms ($q(X)$) for $C_{53}X_{(n\%)}$ systems (X = N or P). In case of P-doped graphene, total magnetizations are given in the last column.

| Strain $n$ / % | N-doped graphene ($C_{53}N_{(n\%)}$) | | P-doped graphene ($C_{53}P_{(n\%)}$) | | |
|---|---|---|---|---|---|
| | $d$(C−N) / Å | $q$(N) / e | $d$(C−P) / Å | $q$(P) / e | $M_{tot}$ / $\mu_B$ |
| −5 | 1.357 | 4.971 | 1.798 | 4.282 | 0.79 |
| −3 | 1.377 | 4.971 | 1.784 | 4.259 | 0.92 |
| −1 | 1.398 | 4.976 | 1.771 | 4.236 | 1.00 |
| 0 | 1.414 | 4.982 | 1.764 | 4.218 | 0.97 |
| 1 | 1.429 | 4.989 | 1.759 | 4.197 | 0.88 |
| 3 | 1.460 | 5.002 | 1.752 | 4.135 | 0.00 |
| 5 | 1.492 | 5.016 | 1.754 | 4.064 | 0.00 |